\documentclass[preprint,amsmath,amssymb,aps,superscriptaddress,12pt,floatfix,linenumbers,showkeys]{revtex4-1}
\usepackage[title]{appendix}
\usepackage[utf8]{inputenc}
\usepackage[T1]{fontenc}
\usepackage{helvet}

\usepackage{graphicx}
\usepackage{capt-of}
\usepackage{geometry}
\usepackage{xcolor}
\usepackage{amsmath}
\usepackage{placeins}
\usepackage{array}

\usepackage{tabularray}
\usepackage{tabularx}
\usepackage{booktabs}
\usepackage{amsfonts}
 
\usepackage{makecell}

\usepackage{setspace}

\usepackage{url}

\usepackage{algorithm2e}
\usepackage{hyperref}

\begin{document}
\nolinenumbers

\title{A microsimulation model of behaviour change calibrated to reversal learning data}
\thanks{Published as: Delos Reyes, R., Lyons Keenan, H., \& Zachreson, C. (2025). A Microsimulation Model of Behaviour Change Calibrated to Reversal Learning Data. \textit{Journal of Artificial Societies \& Social Simulation}, 28(3).}

\author{Roben {Delos Reyes}}
\affiliation{School of Computing and Information Systems, The University of Melbourne, Parkville, Victoria, Australia}

\author{Hugo {Lyons Keenan}}
\affiliation{School of Computing and Information Systems, The University of Melbourne, Parkville, Victoria, Australia}

\author{Cameron Zachreson}
\email{cameron.zachreson@unimelb.edu.au}
\affiliation{School of Computing and Information Systems, The University of Melbourne, Parkville, Victoria, Australia}

\begin{abstract}
Behaviour change lies at the heart of many observable collective phenomena such as the transmission and control of infectious diseases, adoption of public health policies, and migration of animals to new habitats. Representing the process of individual behaviour change in computer simulations of these phenomena remains an open challenge. Often, computational models use phenomenological implementations with limited support from behavioural data. Without a strong connection to observable quantities, such models have limited utility for simulating observed and counterfactual scenarios of emergent phenomena because they cannot be validated or calibrated. Here, we present a simple stochastic microsimulation model of reversal learning that captures fundamental properties of individual behaviour change, namely, the capacity to learn based on accumulated reward signals, and the transient persistence of learned behaviour after rewards are removed or altered. The model has only two parameters, and we use approximate Bayesian computation to demonstrate that they are fully identifiable from empirical reversal learning time series data. Finally, we demonstrate how the model can be extended to account for the increased complexity of behavioural dynamics over longer time scales involving fluctuating stimuli. This work is a step towards the development and evaluation of fully identifiable individual-level behaviour change models that can function as validated submodels for complex simulations of collective behaviour change.
\end{abstract}


\maketitle

\section{Introduction} \label{sec:Introduction}
Behavioural changes are fundamental to many real-world phenomena we seek to understand or predict through computer simulations. Such computational models describe how the actions of individuals in a system change in response to events caused by internal or external factors. Variation in behaviour over time affects the dynamics of the modelled real-world systems and the phenomena that emerge from them, from how infectious diseases spread in a population to how animals adapt to new habitats \cite{Verelst2016, Weston2018, Bauer2013, Leimar2024}. However, in many computer simulations, these individual-level submodels are based on strong assumptions about the underlying factors that influence behaviour and the conditions that drive behaviour change, with limited empirical support. Hence, such models cannot be validated or calibrated based on behavioural data, raising concerns about the verifiability of their results \cite{Verelst2016, Weston2018}. There remains a need for explicit computational representations of behaviour change that are supported by quantitative empirical evidence while maintaining the mechanistic adequacy to simulate observed and counterfactual scenarios. To help address this need, we present a simple (two-parameter) stochastic model of individual behaviour change and demonstrate that its parameters are identifiable from empirical time series data from reversal learning experiments.

To introduce our model, we begin with the general premise that individual decisions about behaviour map available information (stimuli) to a set of actions (responses). As a motivating example, consider endogenous protective behaviours during an infectious disease outbreak: The choice of some individuals to adopt social distancing practices depends on external information from their environment which could be globally available from (e.g.) government and media reports of disease statistics \cite{Collinson2015, Xiao2015}. Additional local information may also be available from communication and observations involving neighbours, friends, and family \cite{Andrews2015}. Such external information can support or contradict information internal to individuals, such as their perceived risks and benefits. How individuals then incorporate this information into their decision-making process may vary from person to person. Such variation often correlates with observable factors such as age, occupation, or socioeconomic status, which may reflect different capabilities and considerations that individuals have when making decisions \cite{Gozzi2021, Gauvin2021, Zachreson2021}. Thus, when modelling behaviour in computer simulations, design choices are often made about the types of information individuals can use to make decisions and the mechanisms by which they translate that information into actions (see, e.g., \cite{Durham2012}). 

In behaviour change models, environmental stimulus could be the number of infected people in an infectious disease model \cite{Poletti2011, Durham2012, Bayham2015}, the physical features of prey in an animal foraging model \cite{Dyer2014, Leimar2024}, or a reward indicating the desirability of a chosen action from the perspective of a global observer with complete knowledge \cite{Xia2021, Le2023}. In simulations, there are two general approaches to modelling how individuals act on this information: rule-based or learning-based. Rule-based approaches condition an individual's behaviour on external information based on predefined threshold criteria or logical relationships \cite{Durham2012, Bayham2015}. In contrast, learning-based approaches assume that individuals are rational and behave optimally by choosing actions that maximize an objective function. The objective functions can be elaborate and may incorporate computing the costs and benefits of multiple strategies or examining past experiences \cite{Poletti2011, Dyer2014}. 

All studies in which such models are developed must make assumptions on the factors that drive behaviour change, but many do so without being calibrated or validated from empirical data \cite{Verelst2016, Weston2018}. In those studies that use behavioural data, the information is often in the form of surveys which are costly to conduct and typically quantify behavioural intentions rather than actual behaviours. Additionally, most previous modelling studies do not draw from the rich body of literature that examines and describes the psychology behind behaviour change which has the potential to provide a stronger theoretical foundation for computational models \cite{Weston2020}. By designing a computational representation of behaviour that is informed by both experimental data and behavioural change theories, we can improve the validity, reliability, and practicality of computer simulations.

\begin{figure}[!t]
    \centering
    \includegraphics[scale=0.425]{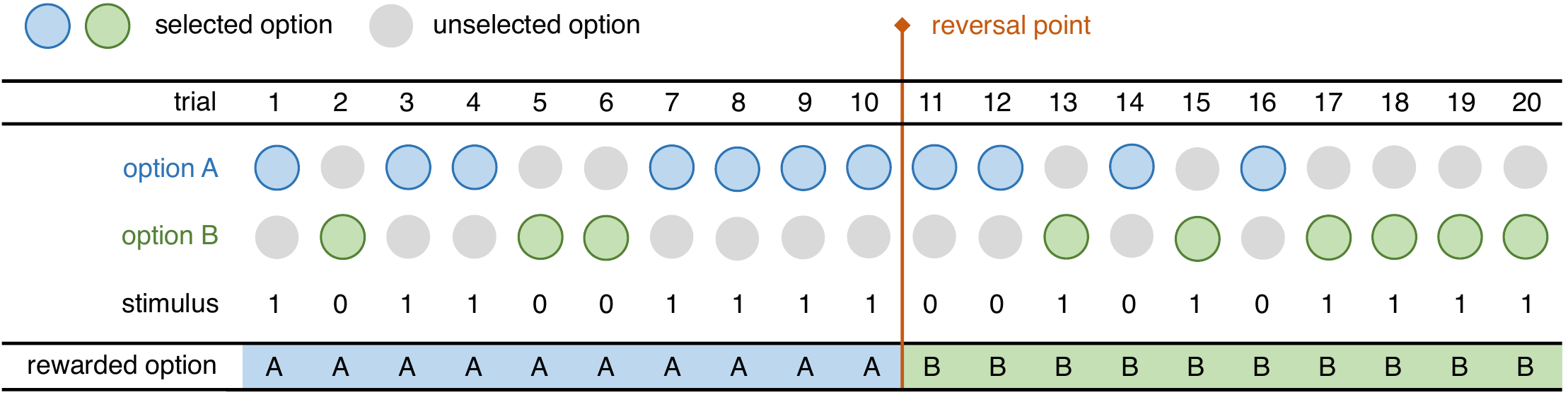}
    \caption{General setup for a reversal learning experiment. An individual selects between two options: A and B. Initially, one option is associated with a positive stimulus and the other option with no stimulus. After the reversal point, this reward association switches. The experiment tests the individual's ability to adapt its behaviour to the reversal of the reward signal.} 
    \label{fig:fig1}
\end{figure}

To develop our microsimulation model of behaviour change, we aim to simulate the reversal learning experiment that is often conducted in neuroscience, psychology, and ecology to study how individuals adapt their behaviour to changes in their environment. An overview of the reversal learning experiment setup is shown in Figure \ref{fig:fig1}. In a typical reversal learning experiment, an individual is tasked to choose between two options: a rewarded and a non-rewarded option. The experiment is repeated in a series of trials and examines whether the individual can learn which of the two options is rewarded. After some time, the reward association is reversed, but the individual is unaware of this reversal. The individual's response to this change in stimulus is then recorded over a second set of repeated trials over which the new reward association may be learned \cite{Rayburn-reeves2011, Dyer2014, Hassett2017, Xia2021, Le2023, Leimar2024}. 

By simulating reversal learning, we developed a microsimulation model of behaviour change with parameters that are fully identifiable from actual behavioural data. To identify the values of these parameters, we used approximate Bayesian computation which is an established approach for calibrating individual-level models to empirical data \cite{Vandervaart2015, Grazzini2017}. We show that this model satisfies three characteristics of a good behavioural change model: (1) mechanistic adequacy, (2) tangibility, and (3) tractability. By mechanistic adequacy, we refer to the degree to which the model's underlying processes are relatable to the features of the behaviour change process (in this case those characterising reversal learning). Tangibility means the model is tangible: its parameters have intuitive meaning and it produces output that can be directly compared to observable quantities or patterns. Tractability means the model is tractable: its parameters are identifiable based on comparison with experiments and it can be calibrated efficiently in practice with sufficiently low computational cost. In addition, we explain how our microsimulation model relates to the COM-B model of behaviour change, a theoretical framework describing that capability (C), opportunity (O), and motivation (M) are the three components necessary for a behaviour (B) to occur.

Below, we provide a detailed description of our agent-based model and reversal learning simulations (Methods, Section \ref{sec:Methods}). We then present the calibration and validation results (Results, Section \ref{sec:Results}), followed by a discussion of the implications of our study (Discussion, Section \ref{sec:Discussion}). We conclude by giving a summary of our findings (Conclusion, Section \ref{sec:Conclusion}). 

\section{Methods} \label{sec:Methods}
In this section, we discuss how our microsimulation model represents reversal learning behaviour. We then describe the experimental data we used to understand and evaluate this behaviour and the experiments that generated the data. Finally, we describe how calibration was performed.

\subsection{Model description}
\begin{figure}[t!]
    \centering
    \includegraphics[scale=0.6]{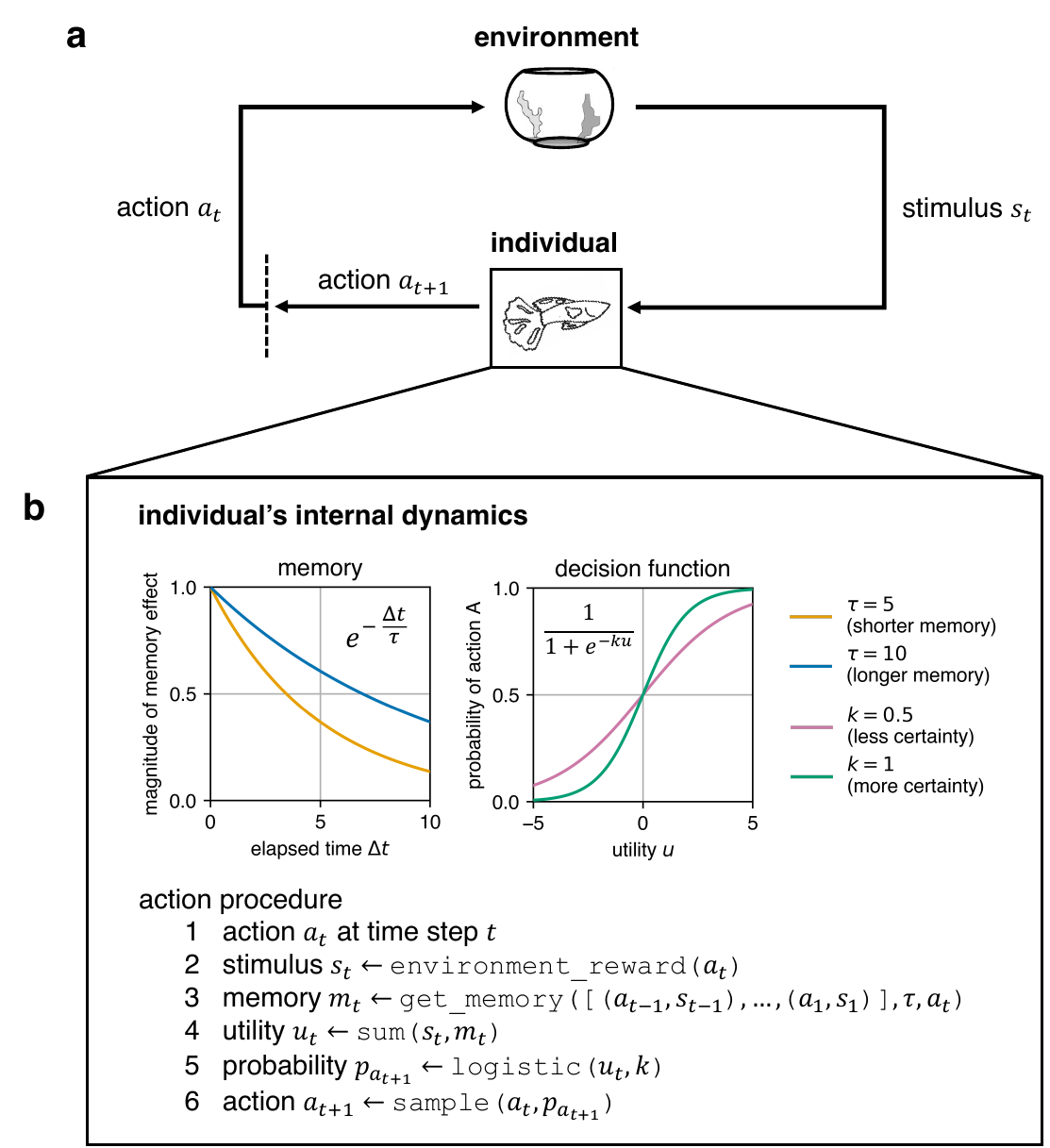}
    \caption{Schematic of the decision-making process in the microsimulation model. (a) The individual chooses an action $a_t$ at time step $t$ and receives a corresponding stimulus $s_t$ from the environment. (b) In choosing the next action $a_{t+1}$, the individual considers the utility $u_t$ of its current action $a_t$. This utility is the sum of the instantaneous stimulus $s_t$ from the environment and the individual's internal memory $m_t$ of past actions and stimuli with a timescale determined by the memory decay $\tau$. The individual decides the next action $a_{t+1}$ using a logistic function with a steepness $k$, which indicates the individual's certainty on the informational value of any given utility.}
    \label{fig:fig2}
\end{figure}

We developed a simple microsimulation model of behaviour change where an individual's behaviour mimics that which is observed in reversal learning experiments. An overview of our microsimulation model is shown in Figure \ref{fig:fig2}. Abstractly, individuals represent any entity capable of making decisions that could produce reward signals. In the model, the individual decides on an action $a$ from a given set of possible actions at every iteration of the simulation. Here, we consider only binary choices for simplicity, but the model can be extended to relax this constraint (see Discussion). 

We created the individual to be capable of sensing its environment and remembering its past experiences which enable it to acquire and learn information for deciding which action to choose, resembling decision-making entities in the real world. Suppose that at the current time step $t$, the individual chooses action $a_t$. The information from the environment comes in the form of a stimulus $s_t$ that encapsulates the impact of the individual's action $a_t$ on the environment. How the environment provides this stimulus $s_t$ depends on the specific context the model is used for. In the context of reversal learning experiments, we set it up as:
\begin{equation}
    s_t =
    \begin{cases}
        +1 &\text{if } a_t \text{ is the rewarded action}\\
        0 &\text{if } a_t \text{ is the non-rewarded action.}
    \end{cases}
\end{equation}
In every simulation of the reversal learning experiment, the environment periodically switches the stimulus associated with the individual's actions and observes the individual's probability of choosing the rewarded actions over time.

The information internal to the individual is encoded as a memory value $m_t$ that quantifies the desirability of an action $a_t$ based on the individual's past experiences. We assume that the individual's past actions and stimuli have an exponentially decaying influence on the individual's memory. If we consider some previous time step $i = t - \Delta t$, the influence $w_i$ of the experience at the previous time step $i$ to the memory $m_t$ at the current time step $t$ is expressed as:
\begin{equation}
    \label{eq:influence}
    w_i = 
    \begin{cases}
        s_{i}e^{-\Delta t / \tau} &\text{if } a_i = a_t\\
        -s_{i}e^{-\Delta t / \tau} &\text{if } a_i \neq a_t
    \end{cases}
\end{equation}
where $a_t$ is the individual's action at the current time step $t$, $a_i$ is the individual's action at the previous time step $i$, $s_i$ is the corresponding stimulus that the individual received for action $a_i$, $\Delta t$ is the elapsed time between time steps $t$ and $i$, and $\tau$ is the timescale of memory. Note that the sign of the influence $w_i$ in Equation {\ref{eq:influence}} is relative to the current action $a_t$. Hence, positive values influence the individual to \textit{stay} with the current action $a_t$ while negative values influence the individual to \textit{change} the current action $a_t$ in the next time step $t+1$. The memory $m_t$ of all past actions and stimuli is then computed as:
\begin{equation}
    \label{eq:memory}
    m_t = \sum_{i=1}^{t-1}{w_i}.
\end{equation}
We note that the current experience at time step $t$ is excluded from the memory calculation to enable the instantaneous stimulus $s_t$ to be weighed differently from the memory $m_t$ of past actions and stimuli.

Together, the external stimulus $s_t$ from the environment and the internal memory $m_t$ of the individual describe the utility $u_t$ of the individual's current action $a_t$ at the current time step $t$:
\begin{equation}
    u_t = s_t + m_t.
\end{equation}
This utility $u_t$ is the combination of all factors that the individual considers when making a decision. Here, we expressed the utility $u_t$ as a sum of the instantaneous stimulus $s_t$ and the internal memory $m_t$ to keep the utility calculation simple. However, other combinations of these two factors can also be considered. Likewise, the model can be extended to account for additional factors that are relevant to an individual's decision-making process.

The individual chooses its next action $a_{t+1}$ at the next time step $t+1$ based on the utility $u_t$. The utility $u_t$ is passed to a logistic decision function:
\begin{equation}
    \label{eq:probability}
    p_{a_{t+1}} = \frac{1}{1 + e^{-k(u_t-b)}}
\end{equation}
where the steepness $k$ indicates the individual's certainty on the informational value of any given utility and $b$ is the midpoint of the function which we set to 0. The logistic decision function returns a probability $p_{a_{t+1}}$ indicating the individual's likelihood of staying with the current action $a_t$ in the next time step $t+1$. Hence, the individual's next action $a_{t+1}$ will be the same as the current action $a_t$ with probability $p_{a_{t+1}}$ or will be the other action with probability $1-p_{a_{t+1}}$.

The individual's behavioural tendencies are characterised by two parameters. The first parameter is the memory decay $\tau$ which specifies the timescale of the individual's memory and hence determines how much information the individual can recall from its past experiences. A higher $\tau$ indicates longer memory, while a smaller $\tau$ means shorter memory. The second parameter that characterises the individual's behavioural tendencies is the logistic decision function's steepness $k$ which indicates the individual's certainty about the value of the information it has accumulated. Given a piece of information, an individual with higher $k$ is more certain about its value, while an individual with lower $k$ is less certain.

\subsection{Relationship to COM-B model of behaviour change}
In designing the components of individual behaviour in our microsimulation model, we used as a basis the components of the COM-B model, a behavioural change theory that has been used to design and implement real-world policies to drive positive changes in people's health behaviour \cite{Michie2011}. The COM-B model proposes that capability (C), opportunity (O), and motivation (M) are the three components that influence an individual's behaviour (B). Capability refers to an individual's psychological and physical abilities to enact a behaviour. Opportunity refers to the external factors that make it possible to enact the behaviour. Motivation refers to the cognitive processes involved in enacting this behaviour. Each component of our microsimulation model was motivated to provide a minimalistic representation of these three COM-B components. 

In any computational model, there has to be an explicit mathematical formulation of how an individual makes decisions for behaviour to be specified. Hence, the individual must have a decision function. Without this decision function, then there is no behaviour. Here, we used a logistic function as the decision function, which by design, naturally represents the individual's capability (C) to choose between two actions. While other formulations of the decision function can be used, the logistic decision function provides a simple mechanism for incorporating the capability of an individual to choose an action based on the available information (in the form of the utility). It also allows an infinite span of behaviours to be specified based on how the logistic decision function's steepness $k$ is selected, where a higher $k$ indicates that an individual is more capable of discerning the value of information and a lower $k$ means that the individual is less capable. Hence, the logistic decision function not only conceptualises the capability of an individual in principle but also characterises the degree of this capability. The utility $u_t$ passed to this logistic decision function represents a form of motivation (M). This utility captures the combined effect of instantaneous reward signals (via the stimulus $s_t$) and reflective evaluations of past actions (via the memory $m_t$). With higher utility values, the individual becomes more motivated to choose one action over the other. Opportunity (O) is represented in the experimental setup where the reversal of reward signals provides an opportunity for the individual to change its behaviour, as mediated by the environment. Should barriers to learning this desired behaviour be included in the experiment, the way the environment provides the stimulus $s_t$ can be changed or the logistic decision function's midpoint $b$ can be biased to make it easier or harder for an individual to change its behaviour.

\subsection{Reversal learning experiments}
\begin{figure}[t!]
    \centering
    \includegraphics[scale=0.7]{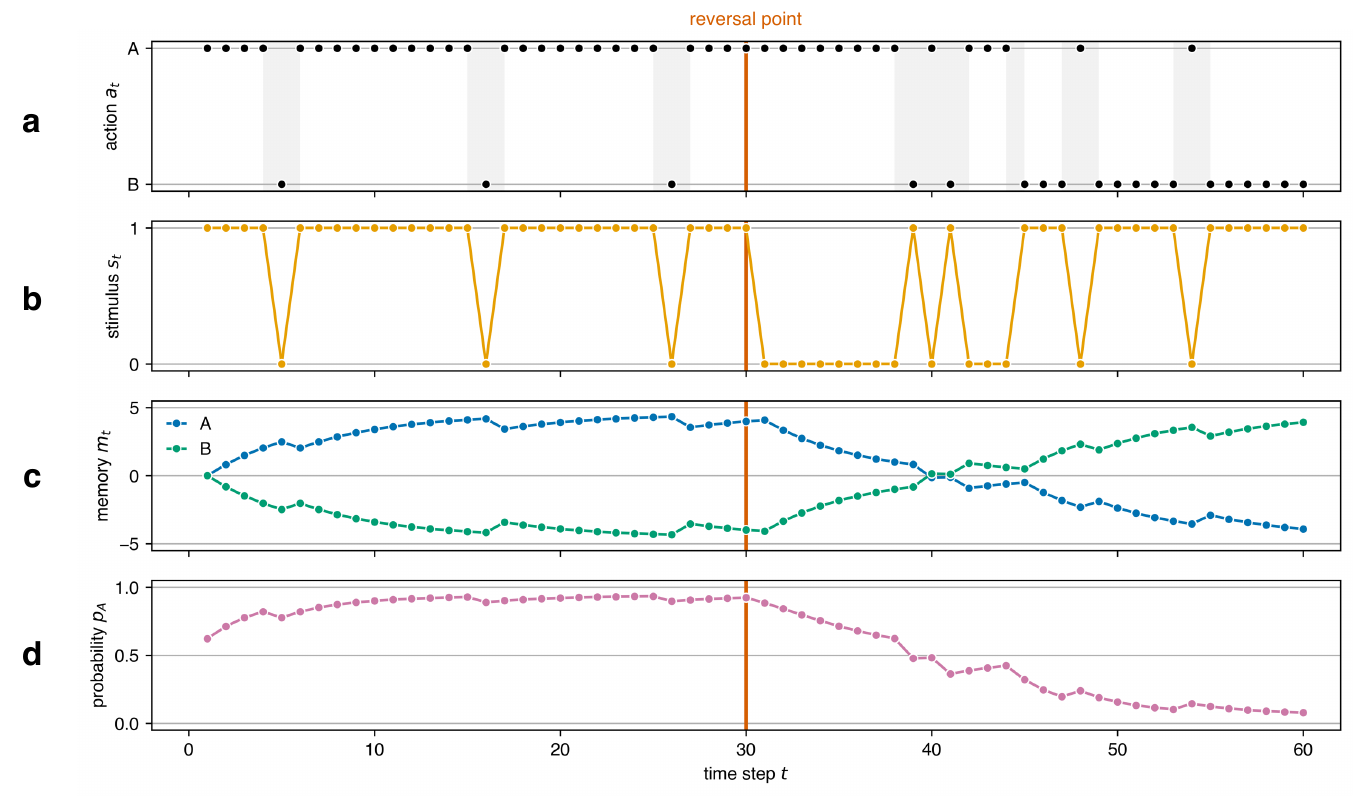}
    \caption{Model dynamics of one simulation run of one individual with a memory decay $\tau=5$ and a logistic function steepness $k=0.5$. (a) The individual selects an action $a_t$ (either $A$ or $B$) at every time step $t$. Highlighted in grey are instances when the individual switches actions. (b) The individual initially receives a stimulus $s_t = +1$ for action $A$ and a stimulus $s_t = 0$ for action $B$ before the reversal point, but this reward association switches after the reversal. (c) The individual has a memory $m_t$ indicating the individual's preference for each action over time. (d) The individual chooses action $A$ with probability $p_A$ based on the stimulus $s_t$ and memory $m_t$.}
    \label{fig:fig3}
\end{figure}

We examined the individual's behaviour using a simulation of single and serial reversal learning experiments. We followed the experimental setup in \cite{Buechel2018} and \cite{Boussard2020} and used their empirical data to calibrate and validate our model. In the single reversal learning experiment on guppies in \cite{Buechel2018}, there were two trial periods with 30 trials in the first trial period and 66 trials in the second trial period. Here, we set the second trial period to 30 for uniformity as done similarly in \cite{Boussard2020}. We note that we used their empirical data on the large-brained guppies only to make a 1-to-1 comparison with the results they presented in the paper. At every trial, the individual (or the guppy) was given two discs to choose from. During the first trial period, one of the discs had a food reward underneath while the other had none, but during the second trial period, the reward associated with each disc was reversed. We simulated this empirical setup by giving the individual in our model two actions to choose from: A and B. For the first trial period, we set the stimulus $s$ to +1 for action A, signifying the presence of the food reward, and 0 for action B, indicating the absence of the food reward. We then switched the stimulus associated with each action after the reversal point to mimic the reward reversal. The model dynamics of one simulation run of this experiment are illustrated in Figure \ref{fig:fig3}. We also simulated a serial reversal learning experiment which extends the single reversal to a series of reversals as conducted in \cite{Boussard2020}. The configurations of the serial reversal learning experiment were similar to the single reversal learning but instead of having only two trial periods separated by one reversal, there were 11 trial periods with ten reversals. These microsimulation model parameters are summarized in Table \ref{tab:table1}. Similar to \cite{Buechel2018} and \cite{Boussard2020}, we report the success rate or the fraction of individuals that successfully chose the rewarded disc at every trial. To validate the results, we compare the success rate in the simulations ($N=100$) with the success rate in the empirical single ($N=24$) and serial ($N=48$) reversal learning experiments. \label{sec:RLsetup}

\subsection{Extended model for serial reversal learning} 
The result of the serial reversal learning experiments in \cite{Boussard2020} revealed a notable characteristic in the individuals' behaviour wherein the individuals' success rate at the end of each trial period decreases over reversals. In our initial simulation runs of the serial reversal learning, we found that our simple two-parameter model was inadequate in capturing this observed phenomenon. Hence, we introduced a third parameter $\theta$ that controls the decay per reversal of the steepness $k$ of the individual's decision function. Specifically, we decreased the value of $k$ by a factor of $\theta \in [0, 1]$ after every reversal period $j$:
\begin{equation}
    k_{j+1} = k_j(1-\theta).
\end{equation}
We can interpret this decay as reducing the impact of information on the individual's decision by decreasing the steepness of its decision function, which means that higher values of utility $u$ are needed to produce the same probability $p_a$. \label{sec:SRLmodel}

\subsection{Calibration: Approximate Bayesian computation}
Similar to previous works, we used approximate Bayesian computation (ABC) to calibrate our microsimulation model to empirical data \cite{Vandervaart2015, Grazzini2017}. ABC rejection sampling is a method for estimating the model parameters' posterior distributions using repeated simulations. Initially, the range of values that each parameter can have is assumed to follow some prior distribution. An ensemble of $N$ models with the parameter values sampled from those priors is then run and the summary statistics of the ensemble are compared with those of the empirical data based on a distance metric. The sampled parameter values are accepted if the distance between the two summary statistics is within the prespecified error threshold $\delta$; otherwise, they are rejected. New parameter values are sampled from the prior distributions and the process repeats until $l$ samples are accepted. These accepted samples characterise the posterior distributions of the model parameters which can be utilised to simulate the observed behaviours from the empirical data. Here, we assumed a uniform prior distribution for each parameter, used the success rate over trials as the summary statistics, and chose the mean squared error as the distance metric. We set the ensemble count $N$ to 100, the error threshold $\delta$ to 0.01, and the sample count $l$ to 1,000. A summary of our ABC parameters is listed in Table \ref{tab:table1}.

\begin{table}[t!]
    \centering
    \footnotesize
    \begin{tabular}{@{\kern0.5em}p{0.17\linewidth}@{}p{0.43\linewidth}@{}p{0.2\linewidth}@{}p{0.18\linewidth}@{}}
    \toprule[.8pt]
    \textbf{Symbol} &\textbf{Description} &\textbf{Value\newline (Single reversal)} &\textbf{Value\newline (Serial reversal)}\\
    \midrule[.6pt]
    \multicolumn{4}{@{\kern0.5em}l@{}}{\textit{Microsimulation Model}}\\
    \midrule[.6pt]
    $n$ &Number of trial periods &2 &11\\
    $r$ &Number of trials per trial period &30 &30\\
    $\tau$ &Memory decay &See Figure \ref{fig:fig5}b &See Figure \ref{fig:fig6}b\\
    $k$ &Steepness of the logistic decision function &See Figure \ref{fig:fig5}b &See Figure \ref{fig:fig6}b\\
    $\theta$ &Decay factor applied to $k$ per trial period &0 &See Figure \ref{fig:fig6}d\\
    \midrule
    \multicolumn{4}{@{\kern0.5em}l@{}}{\textit{Approximate Bayesian Computation}}\\
    \midrule
    $N$ &Ensemble count &100 &100\\
    $\delta$ &Error threshold &0.01 &0.01\\
    $l$ &Sample count &1,000 &1,000\\
    $U(\tau_{min}, \tau_{max})$ &Prior distribution of $\tau$ &$U(0, 50)$ &$U(0, 50)$\\
    $U(k_{min}, k_{max})$ &Prior distribution of $k$ &$U(0, 1)$ &$U(0, 1)$\\
    $U(\theta_{min}, \theta_{max})$ &Prior distribution of $\theta$ &$U(0, 0)$ &$U(0, 1)$\\
    \bottomrule[.8pt]
    \end{tabular}
    \caption{Parameters of the microsimulation model and the approximate Bayesian computation used for the single and serial reversal learning simulations.}
    \label{tab:table1}
\end{table}

For the serial reversal learning simulations, the calibration was performed in two stages. The first stage of the calibration process is identical to the one we used for the single reversal learning simulations where the posterior distributions of the memory decay $\tau$ and the steepness $k$ are calculated for the first two trial periods (separated by a single reversal). In the second stage, we computed the posterior distribution of $\theta$ that captures the behaviour across multiple reversals, with the $\tau$ and $k$ parameter values fixed at the maximum a posteriori (MAP) parameter estimate obtained from the first stage. The MAP parameter estimate specifies the mode of the ABC posterior distributions, representing the most likely value of the parameters given the data. We note that this two-stage calibration procedure is not guaranteed to converge to a maximum likelihood posterior over the joint distribution of $k$, $\tau$, and $\theta$, but nevertheless performs well enough to allow adequate reproduction of the observations after calibration.

\section{Results} \label{sec:Results}
Here, we present the posterior distributions of the model parameters obtained from the calibration process using ABC and compare the simulated behaviours given those parameters with real behaviours observed from empirical reversal learning experiments. 

\subsection{Synthetic reversal learning}
To first demonstrate that the ABC algorithm is able to identify ground-truth parameter values, we performed ABC on a synthetic single reversal learning experiment where $\tau=5$ and $k=0.5$. Using these $\tau$ and $k$ parameter values, we ran 24 simulations (representing the 24 large-brained guppies in the single reversal learning experiment in \cite{Buechel2018}) and calculated their success rate which is shown in a solid black line in Figure \ref{fig:fig4}a. We then executed ABC with a configuration similar to that listed in Table \ref{tab:table1} for single reversal learning. As exhibited in Figure \ref{fig:fig4}b, the posterior distribution identified by the ABC encapsulates the true parameter values of $\tau=5$ and $k=0.5$ (marked in black), and the MAP parameter estimate is close to these parameter values where $\tau=5.1$ and $k=0.56$ (marked in orange). The success rate trajectory from this calibrated model also fits the success rate trajectory from the synthetic data well as shown in Figure \ref{fig:fig4}a, with a mean squared error of 0.006. On the other hand, Figure \ref{fig:fig4} also illustrates that uncalibrated models with $\tau$ and $k$ parameter values outside the ABC posterior distribution show success rate trajectories that do not fit the success rate trajectory from the synthetic data, with mean squared errors ranging from 0.015 to 0.079.

\subsection{Single reversal learning}
\begin{figure}[t!]
    \centering
    \includegraphics[scale=0.55]{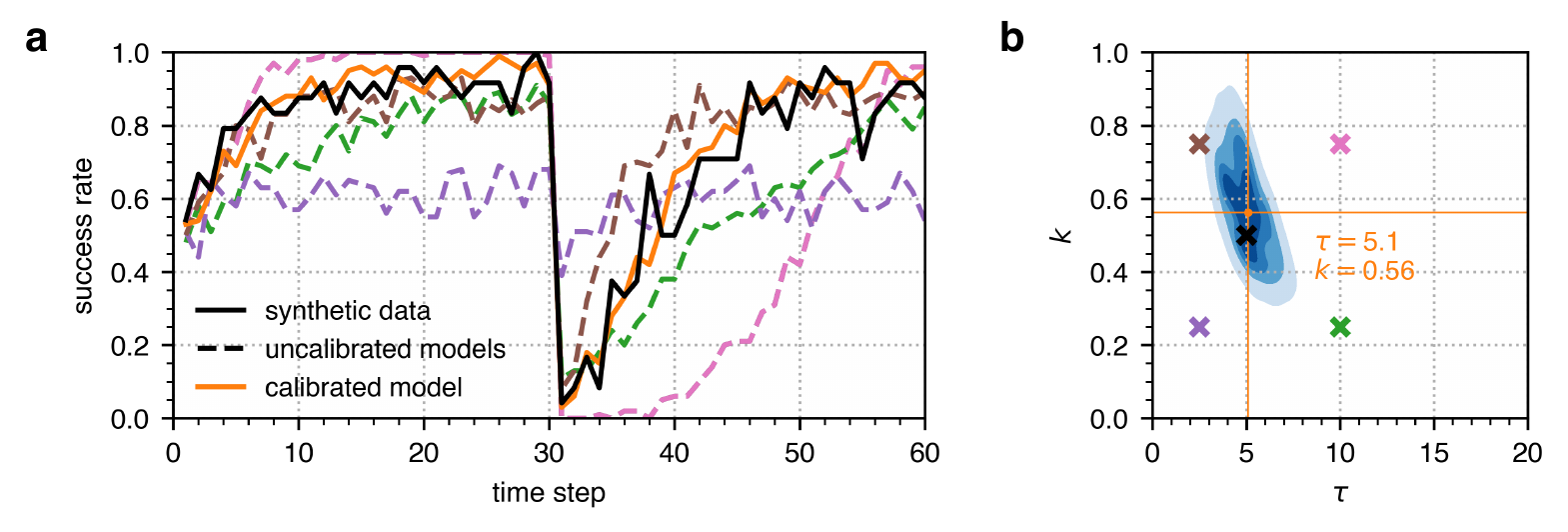}
    \caption{Model calibration to a synthetic single reversal learning experiment with $\tau=5$ and $k=0.5$. The trajectories in (a) show the success rate when using the $\tau$ and $k$ parameter values marked in (b), where the black mark indicates the ground-truth parameter values, the orange mark indicates the MAP parameter estimate, and the other coloured marks indicate parameter values outside the ABC posterior distribution.}
    \label{fig:fig4}
\end{figure}

The simulations recovered the observed trajectory of the success rate given parameter values drawn from the posterior distributions of the memory decay $\tau$ and steepness $k$ obtained from the ABC as shown in Figure \ref{fig:fig4}. The success rate shows the average success across 100 individuals with the same $\tau=9.7$ and $k=0.26$ parameters. At the start of the first trial period, the success rate is only about 50\% because the individuals are still unaware of the task at hand. The individuals eventually learn which option has the reward after about 20 trials wherein the success rate reaches 80\%, which is the criterion set in \cite{Buechel2018} to indicate that an individual learned to discriminate the rewarded and non-rewarded options. When the reward is reversed after the first trial period, the success rate drops from about 90\% to 10\%. This drop in performance indicates the effect of memory on the individuals' decision-making process. Despite the change in the stimulus from the environment, the individuals' memory still favours the previously rewarded option. The individuals eventually learn the new reward association over time such that the success rate steadily increases back to 80\%. The mean squared error between the observed and simulated success rate trajectories is 0.007.

\begin{figure}[t!]
    \centering
    \includegraphics[scale=0.55]{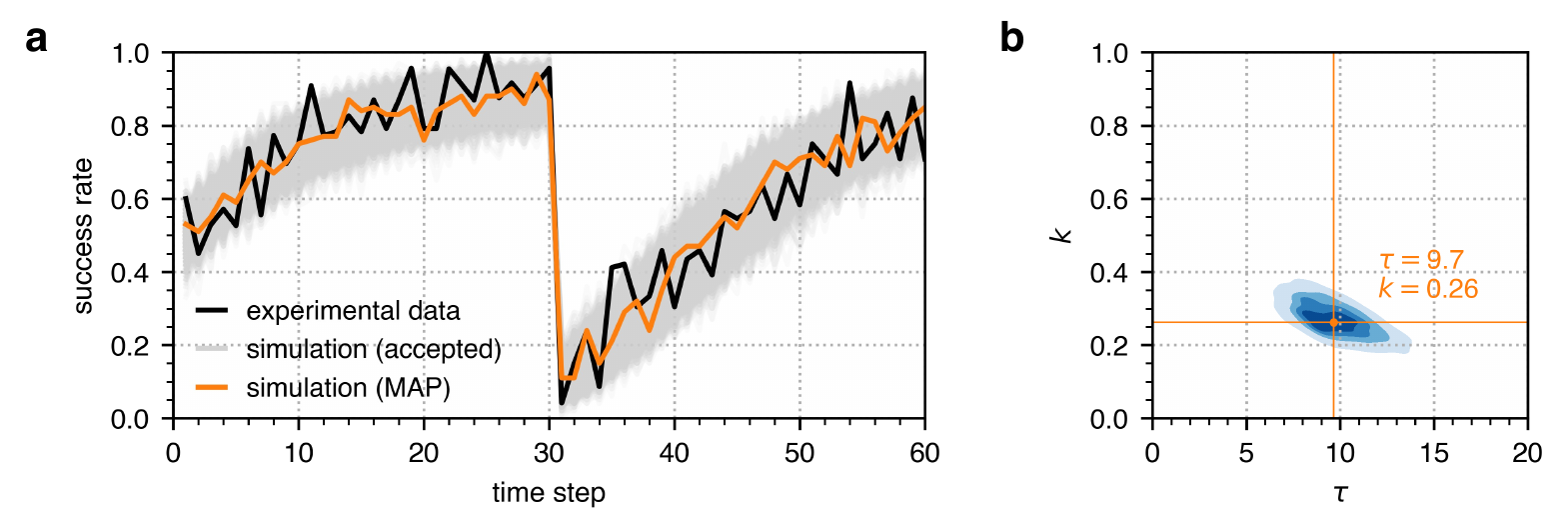}
    \caption{Model calibration to single reversal learning experiments. The trajectory in (a) shows the success rate over an ensemble of 100 individuals in a single reversal learning experiment in which the reward reversal occurs after 30 trials. The black trace is the experimental data from \cite{Buechel2018}, the grey traces are the simulation results using all accepted ABC samples, and the orange trace is the simulation result using the MAP parameter estimate. The ABC posterior is shown in (b) with the MAP parameter estimate indicated by the orange point.}
    \label{fig:fig5}
\end{figure}

\subsection{Serial reversal learning}
As shown in Figure \ref{fig:fig5}a, the first calibration stage of the serial reversal learning simulations also recovered the observed success rate in the first of ten reversals from the experimental data from \cite{Boussard2020}. The individuals' success rate goes beyond 80\% after several trials, exhibits a huge drop right after the reversal point, and then rises to about 90\% eventually. With $\theta=0.12$ obtained from the posterior distribution of $\theta$ after the second calibration stage, the simulated data captures the decline in success rate in the succeeding reversals as depicted in Figure \ref{fig:fig5}c. The success rate after the first reversal goes back up to about 90\% but only reaches about 70\% after the tenth reversal. The mean squared error between the observed and simulated success rate trajectories is 0.008.

\begin{figure}[t!]
    \centering
    \includegraphics[scale=0.55]{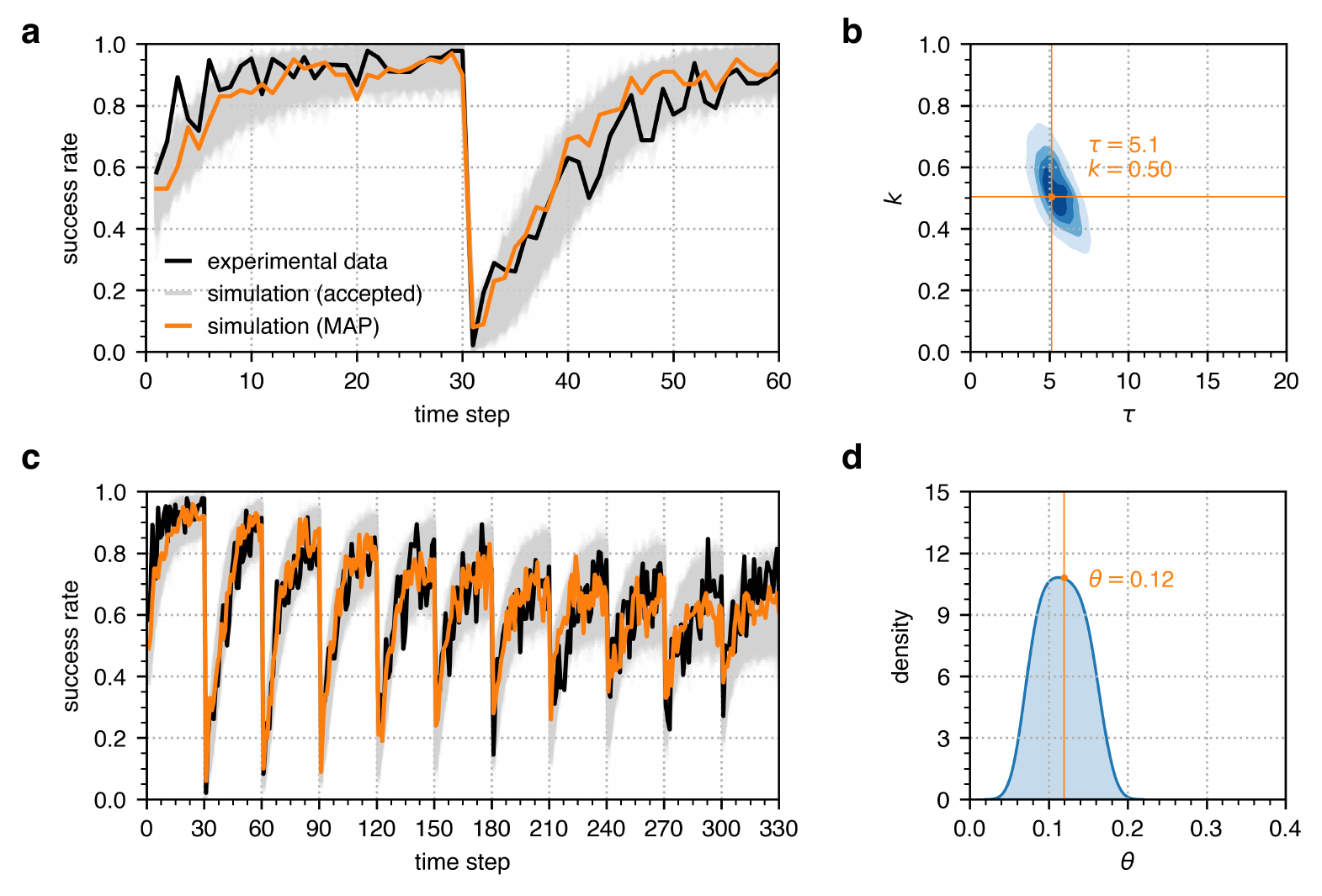}
    \caption{Two-stage model calibration to serial reversal learning experiments. The trajectories in (a) and (c) show the success rate over an ensemble of 100 individuals in the first reversal after the first calibration stage and in all ten reversals of a serial reversal learning experiment after the second calibration stage, respectively. The black traces are the experimental data from \cite{Boussard2020}, the grey traces are the simulation results using all accepted ABC samples, and the orange traces are the simulation results using the MAP parameter estimate. The ABC posterior is shown in (b) for the first calibration stage and (d) for the second calibration stage with the MAP parameter estimate indicated by the orange point.}
    \label{fig:fig6}
\end{figure}

\section{Discussion} \label{sec:Discussion}
Representation of individual behaviour change in computational models of real-world systems is essential to understanding and predicting phenomena emerging from collective behaviour. All these models must make assumptions on how behaviour changes, but many rely on assumptions with limited support from quantitative empirical data which may limit the validity of simulation results. Here, we demonstrated a simple model of behaviour change that can simulate actual behaviours observed in single reversal learning experiments. We further showed that the model can be easily extended to simulate more complex behaviours in serial reversal learning experiments with longer time scales. 

We developed a microsimulation model wherein an individual's behaviour is characterised by three parameters: (1) the memory decay $\tau$ that indicates the extent to which an individual uses past experiences to make decisions, (2) the steepness $k$ of the logistic decision function that determines an individual's certainty on the value of given information, and (3) the factor $\theta$ that specifies the decay of an individual's certainty about information value over multiple reward reversals. We calibrated these model parameters using ABC given the empirical data from \cite{Buechel2018} and \cite{Boussard2020}. As shown in Figures \ref{fig:fig4} and \ref{fig:fig5}, the simulated behaviours using those identified parameters are well-matched to the ensemble statistics observed in the empirical experiments. 

These results indicate that our proposed behavioural change model satisfies mechanistic adequacy, tangibility, and tractability. First, with only two parameters, our model was able to simulate the following qualitative patterns observable in reversal learning data: (1) the capacity to learn associations between decisions and rewards, (2) stochasticity in the decision process, and (3) transient persistence of learned behaviours after the original stimulus is reversed. Further, by adding one additional parameter, our model reproduced the long-term behaviour observed in experiments with multiple reversals. In all cases, calibration to observed behaviour located compact regions in the parameter spaces of our models, indicating that the parameters can be efficiently identified from empirical data. Our work thus demonstrates the feasibility of modelling and simulating real-world behaviours using behavioural data.

Individual behaviours in reversal learning experiments demonstrate a balance between behavioural flexibility (adaptation in response to changing stimuli) and persistence (the tendency for learned behaviour to persist even when rewards are inconsistent). Such a balance is essential for individuals to survive in a dynamically changing world, so it is not surprising that the observed behaviour in guppies in the single reversal learning experiment examined here is consistent with observations of bees \cite{Dyer2014}, birds \cite{Rayburn-reeves2011}, rodents \cite{Le2023}, monkeys \cite{Hassett2017}, and humans \cite{Rayburn-reeves2011, Xia2021}. The decline in success rate observed in the serial reversal experiments we studied suggests a complex mechanism of information processing in which the perceived value of information appears to decay after many fluctuations in the reward signal. During the COVID-19 pandemic where non-pharmaceutical interventions were repeatedly imposed, lifted, and reinstated, a similar decline in adherence to costly behaviours such as physical distancing was observed \cite{Petherick2021}. Designing computational models to have the mechanistic adequacy to simulate such general and specific behavioural change processes could, for example, help better understand animal foraging behaviour \cite{Dyer2014} or enable the design of more effective policy interventions during infectious disease outbreaks \cite{Petherick2021}. However, to be useful, such models need to have tractable computational complexity and cost, while also maintaining tangible sets of observable inputs and outputs.  

In order to calibrate and validate the model to reproduce such real-world behaviours, there is a need for suitable behavioural data. However, because those data are difficult to collect, many existing models rely on assumptions with limited empirical support. Models that do incorporate behavioural data rely mainly on behavioural surveys \cite{Durham2012, Fierro2013, Zhong2013, Bayham2015, Karimi2015}, which are difficult to scale to larger populations and often characterise individuals' behavioural intentions that are not necessarily consistent with their actual behaviours \cite{Verelst2016, Weston2020}. Behavioural data capturing individuals' actual actions provides a more accurate representation of real-world behaviours. For example, mobility data such as GPS trajectories from mobile phones can be readily collected under different scenarios and conditions \cite{Badr2020, Xiong2020, Ilin2021}. Other indicators of behaviour such as television viewing duration \cite{Springborn2015} or home internet usage \cite{Zachreson2021} can also help quantify ensembles of discrete individual choices. Developing models of individual decision making that can be calibrated to such ensemble data is important for ensuring that simulated behaviours are tangible and valid in the context of the collective behaviour of interest such as the endogenous response to infectious disease outbreaks.

Beyond reproducing real-world behaviours, ensuring that the model parameters that generated such behaviours are identifiable from empirical data is also crucial when designing behavioural change models. Because models are often used to inform decision making, it is desirable to have full and explicit information on what drives a model's dynamics to improve its interpretability and reliability. By using ABC to calibrate the model, we can characterise the posterior distributions of the model parameters and demonstrate the uncertainty in the model's dynamics and predictions. Quantifying this model uncertainty could make the analysis and utilisation of models more robust and effective because it provides information on how to interpret their results and on which instances they should be relied on \cite{Edeling2021}.

Previous reviews of behavioural change models have also highlighted that many computational models are not informed by behavioural change theories which provide foundations for understanding, predicting, and influencing individual behaviours. Despite such theories being used in designing and implementing nationwide public health policies and disaster preparedness and response strategies \cite{Michie2011, Angus2013, West2020}, they are rarely incorporated into models that are also developed to support decision making in those scenarios \cite{Weston2020}. We designed behaviour in our microsimulation model based on the COM-B model of behaviour change which states that capability (C), opportunity (O), and motivation (M) are the three key components that influence behaviour (B) \cite{Michie2011}. Our proposed microsimulation model provides a simple computational representation of behaviour whose components can be mapped to those of the COM-B model, and which can be modified or extended to account for the multiple other factors described in psychological theories of behaviour change. \label{sec:comb_model}

Our study has several limitations. The first is that we only considered a binary set of actions, which many decision-making processes can be reduced to. However, the model can be extended to incorporate more than two actions using other decision functions like a softmax function \cite{Sutton2018}. The second limitation is that we used an aggregate behavioural measure to calibrate the model parameters. While those parameters may not be characteristic of each individual's behaviour, our approach could be extended to represent ensembles of individuals with values of individual-level model parameters drawn from population-level hyperdistributions. In practical terms, the approach presented here can be viewed as a technique for building individual-level models that are consistent with ensemble observations. This approach is consistent with typical observables of real-world systems, which are often quantified in aggregate for subpopulations such as communities or demographic strata \cite{Springborn2015, Badr2020, Xiong2020, Ilin2021}.

\section{Conclusion} \label{sec:Conclusion}
In summary, we developed a simple yet extensible microsimulation model of behaviour change that is mechanistically adequate to simulate reversal learning. We calibrated the parameters of our model from empirical behavioural data using approximate Bayesian computation and validated that the model reproduces behaviours observed in single and serial reversal learning experiments given those parameters. This work is a step towards developing computational models of behaviour change whose parameters are fully identifiable from actual behavioural data and are grounded in psychological theories. Such models are essential for understanding and predicting how behavioural changes affect emergent real-world phenomena with better accuracy, reliability, and practicality.

\section*{Code availability}
The code used for this work is available at: \url{https://github.com/rddelosreyes/abm-behaviour-change}


\bibliographystyle{apalike}
\bibliography{refs}

\end{document}